\documentclass[12pt]{iopart}

\usepackage{iopams}

\usepackage[normalem]{ulem}

\usepackage{graphicx,color}
\usepackage{bm}

\usepackage{amsthm}

\def\BraKet#1#2{\langle#1\vert #2\rangle}
\def\Ket#1{\left\vert #1\right\rangle}
\def\Bra#1{\left\langle #1\right\vert}

\def\e{\mathrm{e}}
\def\ii{\mathrm{i}}
\def\d{\mathrm{d}}

\begin{document}
\title{A quantum particle in a box with moving walls}
\author{Sara Di Martino$^{1}$, Fabio Anz\`{a}$^{2}$, Paolo Facchi$^{3,4}$, Andrzej Kossakowski$^{5}$, Giuseppe Marmo$^{6,7}$, Antonino Messina$^{2}$, Benedetto Militello$^{2}$ and Saverio Pascazio$^{3,4}$}
\address{$^{1}$Dipartimento di Matematica, Universit\`a di Bari, I-70125  Bari, Italy}
\address{$^{2}$Dipartimento di Fisica e Chimica, Universit\`a di Palermo, I-90123 Palermo, Italy}
\address{$^{3}$Dipartimento di Fisica and MECENAS, Universit\`a di Bari, I-70126  Bari, Italy}
\address{$^{4}$INFN, Sezione di Bari, I-70126 Bari, Italy}
\address{$^{5}$Institute of Physics, Nicolaus Copernicus University, 87-100 Toru\'n, Poland}
\address{$^{6}$Dipartimento di Fisica and MECENAS, Universit\`a di Napoli ``Federico II", I-80126  Napoli, Italy}
\address{$^{7}$INFN, Sezione di Napoli, I-80126  Napoli, Italy}

\date{\today}

\begin{abstract}
We analyze the non-relativistic problem of a quantum particle that
bounces back and forth between two moving walls. We recast this
problem into the equivalent one of a quantum particle in a fixed box
whose dynamics is governed by an appropriate time-dependent
Schr\"odinger operator.
\end{abstract}


\vspace{2pc}
\maketitle

\section{Introduction}\label{sec:introduction}

In 1949 Enrico Fermi~\cite{Fermi} proposed a model for cosmic ray
production in which a particle moves in inhomogeneous magnetic fields. This problem was reconsidered by 
Ulam~\cite{Ulam}, who introduced the so-called Fermi accelerator in which a ball bounces back and forth between 
two oscillating walls.  He numerically discovered 
that such a system displays a very irregular behavior, and since then it has become a paradigmatic model to illustrate  
the chaotic nature of stochastic
trajectories in systems with two degrees of freedom~\cite{Lichtenberg}. A
non-relativistic quantum version of this problem was first studied, to the best of our knowledge,  by Doescher 
and Rice~\cite{amjphys}, who analyzed the case in which one
of the walls is static while the second one moves with a linear
velocity. A number of articles followed, studying the problem from
different aspects. In particular, the solvability of the model was investigated in a series of papers by Dembi\'{n}ski, Makowski, \textit{et al}~\cite{dembi1,dembi3,dembi4,dembi7} who exhibited several exact and perturbative solutions. Moreover, Dodonov \textit{et al}~\cite{Dodonov} analyzed uniformly moving walls, including the cases of adiabatic motion and almost sudden changes
of the wall position, while Munier \textit{et al}~\cite{Munier} and Konishi and Paffuti~\cite{Konishi} investigated the behaviour of the average energy. The problem in higher dimensions was considered by Lenz \textit{et al}~\cite{Lenz} and by Del Campo and Boshier~\cite{Del_Campo}, who focused on Bose-Einstein condensates and experimental realizations.

Generally speaking, problems of this sort are tackled by searching solutions of the time-dependent Schr\"odinger equation related to the appropriate (time-dependent or static) Hamiltonian satisfying time-dependent boundary conditions. Such a direct approach,
however, overviews the structural difficulty that the problem is
ill-posed in its own, since one is requested to find solutions
belonging to a time-dependent Hilbert space. In other words, the
constraints characterizing the physical problem under scrutiny (no
accessibility into time-dependent regions of the line) are
mathematically taken into account by writing down a dynamical
equation defined within a continuously time-changing space domain.
This is an ill-posed problem since $t$ is among the independent
variables of the same equation. In this direction, a rigorous
analysis of time-dependent Schr\"odinger operators and/or time-dependent domains can be found in the work 
of Yajima~\cite{Yajima87,Yajima96}, Dell'Antonio \textit{et al}~\cite{Dell'Antonio} and Posilicano 
\textit{et al}~\cite{Posilicano07,Posilicano11}.

The quantum bouncer problem is also interesting in the general
context of quantum boundary conditions. Indeed, the behavior of the
wave function at the boundary must reflect the quantum mechanical necessity to define observables in terms of self-adjoint
operators~\cite{von, Reed, QMbook, Galindo}. The most commonly
used boundary conditions in physics are Dirichlet (vanishing wave
function at the boundary), Neumann (vanishing normal derivative),
and periodic ones. The interplay among different boundary conditions, with focus on an interval of the real line, was analyzed in~\cite{AIM, AFMP} and their role and importance has been recently stressed in an interesting 
article~\cite{Wilczek}, where varying boundary conditions are
viewed as a model of spacetime topology change. Notable
applications arise in the context of the Casimir effect and its
dynamical version, giving rise to photon generation in a microwave
cavity with time-dependent boundary conditions~\cite{photons,Para}.

In this paper, we recast the problem of a quantum particle in a
box with moving walls into the problem of a quantum particle
placed in a box with fixed walls and governed by a time-dependent
Schr\"odinger operator. Important properties of the system will be analyzed, such
as the rate of energy change. 
The backbone of the article is the following. 
In section~\ref{sec:model} we introduce the physical problem and the relevant
Hamiltonian model, and introduce the unitary transformation that enables us to recast the original problem into a problem with fixed
boundary conditions. In section~\ref{sec:energy} we derive the
rate equation for the energy of the particle, singling out the
role of contact terms with the walls, expressed by the presence of
time derivatives of the wave function at the positions of the two
walls. In section~\ref{sec:potential} we complete our treatment by
including a potential acting on the particle. Finally, in section~\ref{sec:conclusion} 
we give some concluding remarks and discuss future 
perspectives.

\section{Hamiltonian Model}
\label{sec:model}

\subsection{A particle bouncing between moving walls}

Consider a free massive particle in a one-dimensional box
delimited by two walls at positions $a= -l/2+d$ and $b= l/2+d$.
The box has therefore width $l>0$ and is centered at $d\in\mathbb{R}$, where
$l$ and $d$ are functions of time, $l(t)$ and $d(t)$.

If the walls are still, the dynamics is described
 by the Schr\"odinger equation
 \begin{equation}
\ii \hbar \frac{\d }{\d t} \psi(t)=\frac{p^2}{2m}\psi(t)
\label{eq:Mov_SchrEq}
\end{equation}
in the Hilbert space $L^2(I_{l,d})$ of square integrable functions on the interval
\begin{equation}
I_{l,d}=[a,b]=\left[d-\frac{l}{2},d+\frac{l}{2}\right].
\end{equation}
Here $p^2 = -\hbar^2\frac{\d^2}{\d x^2}$, acting on functions belonging to $\mathcal{H}^2(I_{l,d})$, the Sobolev space of square
integrable functions with square integrable  second derivative, and vanishing at the ends (Dirichlet boundary conditions). Namely,
\begin{equation}
D_{l,d}=\left\{ \psi\in
\mathcal{H}^2\left(I_{l,d}\right), \; \psi\left(d-\frac{l}{2}\right)=\psi\left(d+\frac{l}{2}\right)=0\right\}.
\label{eq:Dld}
\end{equation}
However, when the walls are moving, the interpretation of an apparently innocuous equation as~(\ref{eq:Mov_SchrEq}) requires some care. At first sight, if interpreted, as above, as an equation among different Hilbert spaces, it is meaningless. Indeed, the time derivative
\begin{equation}
\frac{\d }{\d t} \psi(t)=\lim_{\epsilon\to 0} \frac{\psi(t+\epsilon)-\psi(t)}{\epsilon},
\end{equation}
would involve the sum of two vectors that belong in general to different Hilbert spaces $L^2(I_{l(t),d(t)})\neq L^2(I_{l(t+\epsilon),d(t+\epsilon)})$.
Therefore, in order to correctly formulate the problem, one has to give a meaning to the time derivative.

This can be accomplished by embedding $L^2(I_{l,d})$ in the larger Hilbert space of a particle on a line $L^2(\mathbb{R})$ viewed as the direct sum
\begin{equation}
L^2(\mathbb{R})=L^2(I_{l,d})\oplus L^2(I_{l,d}^c),
\label{eq:L2sum}
\end{equation}
where $X^c = \mathbb{R}\setminus X$ denotes the complement of the
set $X$. Thus, the kinetic energy operator in the common enlarged
space is given by the Hamiltonian
\begin{equation}
H_0(l,d)=\frac{p_{l,d}^2}{2m}= -\frac{\hbar^2}{2 m} \frac{\mathrm{d}^2}{\mathrm{d} x^2}\oplus_{l,d} 0,
\label{Ham}
\end{equation}
acting on wave functions of the form $\psi + \phi$ with $\psi\in D_{l,d}$ and $\phi\in
L^2\left( I_{l,d}^c \right)$.
A heuristic derivation of the Hamiltonian~(\ref{Ham}) from the
Hamiltonian of a free particle on a line, with the moving walls
implemented by a quantum Zeno dynamics~\cite{ZenoMP}, is given in
the Appendix.

It is worth noticing that the direct sum decomposition  both
in~(\ref{eq:L2sum}) and in~(\ref{Ham}), in the case of moving
walls, is in fact time-dependent. Therefore, one has to cope with
a Schr\"odinger equation with  a time-dependent Hamiltonian on a
time-dependent domain. The common attack strategy of this kind of
problems is to describe the evolution by unitarily equivalent
Schr\"odinger operators acting on a common fixed domain. We will
also be following this approach in the next sections.

\subsection{Reduction to constant boundary conditions (static
domain)}

The intervals $I_{l,d}$ with $l>0$ can be mapped to a standard
reference interval
\begin{equation}
I=I_{1,0} = \left[-\frac{1}{2},\frac{1}{2}\right],
\end{equation}
via a translation $x \to x -d$ followed by a dilation $x \to x/l$. These transformations of $\mathbb{R}$ can be lifted to transformations on $L^2(\mathbb{R})$, implemented by one-parameter (strongly continuous) unitary groups:
\begin{equation}
(T(d) \psi) (x) = \psi (x-d), \qquad (D(s) \psi ) (x) = \e^{-s/2} \psi(\e^{-s} x).
\end{equation}

Let us therefore consider the unitary transformation
\begin{equation}
\fl \qquad U(l,d):L^2(\mathbb{R}) \rightarrow L^2(\mathbb{R}), \qquad  U(l,d) = D(\ln l)^\dag T(d)^\dag = D(-\ln l) T(-d),
\label{eq:Uld}
\end{equation}
which acts as
\begin{equation}
(U(l,d)\psi)(\xi)=\sqrt{l}\psi(l \xi+d).
\end{equation}
This transformation obviously maps $L^2(I_{l,d}^c)$ onto $L^2(I^c)$, and, more interestingly,  the domain~(\ref{eq:Dld}) onto the
fixed domain $D=U(l,d) D_{l,d}$
\begin{equation}
 D=\left\{
\phi\in \mathcal{H}^2(I),\; 
 \phi\left(-\frac{1}{2}\right)=\phi\left(\frac{1}{2}\right)=0\right\}\subset L^2(I),
\label{eq:Dirichlet}
\end{equation}
describing Dirichlet boundary conditions in a fixed box $I$.
Moreover, the Hamiltonian~(\ref{Ham}) is mapped into
\begin{equation}
H(l) = U H_0 U^{\dagger}= \frac{1}{l^2}
\frac{p^2}{2m} \oplus 0 =-\frac{1}{l^2} \frac{\hbar^2}{2 m}
\frac{\mathrm{d}^2}{\mathrm{d} x^2}\oplus 0,
\label{eq:H(l,d)}
\end{equation}
with \emph{fixed} domain $D\oplus L^2(I^c)$.

It is worth noticing  that the unitary transformation~(\ref{eq:Uld}) yields constant boundary conditions only
 when the latter involve  only the
wave function or only  its derivative. This is the case of
Dirichlet and Neumann boundary conditions but not the Robin ones~\cite{AIM},
\begin{equation}
\psi '\left(d\pm\frac{l}{2}\right)=\alpha\, \psi
\left(d\pm\frac{l}{2}\right).
\end{equation}

Let us now take a closer look at $H(l)$. When one goes from the
interval $I_{l,d}$ to the fixed interval $I$, through the unitary
transformation $U$, the eigenfunctions of $H(l)$ are a rescaled
version of the ones of $H_0$, while the eigenvalues scale by a
factor $l^{-2}$ (the prefactor of $p^2$), since the distance
between the walls change by a factor $l$. This is formally
expressed by the  transformation law of the momentum operator on
the line:
\begin{equation}
U(l,d) p U(l,d)^\dag = D(\ln l)^\dag\, p \, D(\ln l) = \frac{p}{l},
\end{equation}
since the translation group $T(d)=\exp(-\ii d p /\hbar)$ commutes with its generator $p$.

From now on, for the reader's convenience, we will denote the (relevant parts of the) wave functions by $\psi(x)$ with $x\in I_{l,d}$  and by $\phi(\xi)$ with $\xi\in I$, in the frame with moving and fixed walls, respectively.

The Schr\"odinger equation in the frame with fixed box will
contain additional geometrical terms, induced by the time-dependent
transformation $\phi(t)=U(l(t),d(t))\psi(t)$. Taking into account
the Schr\"odinger equation in the original frame
\begin{equation}
\ii \hbar \frac{\d}{\d t}{\psi}(t) = H_0(l(t),d(t)) \psi(t),
\end{equation}
we obtain
\begin{equation}
\ii \hbar \frac{\d}{\d t}{\phi}=\left(H(l)+\ii\hbar
\dot{U}(l,d)U^\dagger(l,d) \right)\phi, \label{eq:newSchrodinger}
\end{equation}
in an appropriate domain.

Let us compute the geometric contribution $\ii \hbar \dot{U}U^\dagger$ step by step. First of all the
action of $\dot{U} = \d U(l(t),d(t))/\d t$ on a test function $\psi$:
\begin{equation}
\fl \qquad \left(\frac{\d U}{\d t} \psi \right)(\xi)=\frac{\d}{\d
t} \left(\sqrt{l} \psi (l\xi+d) \right)= \frac{\dot{l}}{2\sqrt{l}}
\psi (l\xi+d)+\sqrt{l}(\dot{l}\xi+\dot{d}) \psi' (l\xi+d).
\end{equation}
Therefore, being
\begin{equation}
(U^\dagger(l,d) \phi)(x) =   (T(d) D(\ln l)  \phi)(x) = \frac{1}{\sqrt{l}}\,
\phi\left(\frac{x-d}{l}\right),
\end{equation}
we have
\begin{equation}
\ii \hbar \frac{\d U}{\d t} U^\dagger \phi (\xi) =  \ii
\frac{\hbar}{2} \frac{\dot{l}}{l} \phi (\xi) + \ii \hbar  \left( \frac{\dot{l}}{l}\xi+\frac{\dot{d}}{l} \right) \phi'
(\xi),
\end{equation}
that is
\begin{equation}
\ii \hbar \frac{\d U}{\d t} U^\dagger  =  - \frac{\dot{l}}{l} \left(x  p- \ii
\frac{\hbar}{2}\right)  - \frac{\dot{d}}{l}\ p,
\end{equation}
with $x$ and $p$  the standard position and momentum operators. Thus, the geometric generator of the unitary transformation reads
\begin{equation}
K(l,d)= \ii \hbar \frac{\d U}{\d t} U^\dagger= - \frac{\dot{l}}{l}\ x \circ p-\frac{\dot{d}}{l}\ p,
\label{eq:transport}
\end{equation}
where  $A\circ B = (AB+BA)/2$ is the symmetrized (Jordan) product
of the operators $A$ and $B$, and the canonical commutation relation $[x, p]=\ii\hbar$ has been used.

By plugging  expressions~(\ref{eq:H(l,d)})
and~(\ref{eq:transport}) into Eq.~(\ref{eq:newSchrodinger}), we
finally  obtain the expression of the Schr\"odinger equation for the
transformed state  in the reference frame with fixed walls:
\begin{equation}
\label{eq:StaticH_SchrEq}
\ii \hbar \frac{\d }{\d t}\phi=
\Big(H(l)+K(l,d) \Big)\phi
=\left(\frac{1}{l^2}\frac{p^2}{2m
}-\frac{\dot{l}}{l}\ x \circ p-\frac{\dot{d}}{l}\ p\right)\phi\,.
\end{equation}
Apart from the term $p^2/(2ml^2)$ that we discussed above, in
this new Schr\"odinger equation we find two  geometric terms.
Both of them are due to the fact that application of the unitary transformation fixes the walls. This consists on a translation $T(d(t))= \exp(-\ii d(t) p / \hbar)$, generated by the momentum operator $p$ with domain $\mathcal{H}^1(\mathbb{R})$, and a dilation $D(\ln l(t) ) = \exp(-\ii \ln l(t) x\circ p/\hbar)$, generated by the virial operator $x\circ p$  on its maximal domain in $L^2(\mathbb{R})$.
Incidentally, we notice that the emergence of the virial operator in Eq.~(\ref{eq:StaticH_SchrEq}) was recently advocated  by Jarzynski~\cite{Jarzynski} on the basis of symmetry considerations.

Let us finally show that the fixed domain $D$ is a domain of self-adjointness for the total Hamiltonian in~(\ref{eq:StaticH_SchrEq}). Notice that $K(l,d)$ is relatively bounded with $0$ relative bound with respect to $H(l)$, for every $d\in\mathbb{R}$ and $l>0$. Indeed, $p$ is  $p^2$-bounded on $D$ and $\| x p \phi \| \leq \| p \phi\| /2$ for any $\phi \in D$. Thus, since $K(l,d)$ with domain $D$ is a symmetric operator, by the Kato-Rellich theorem the total Hamiltonian $H(l) + K(l,d)$ with domain $D(H(l)+K(l,d))= D(H(l))=D$ is self-adjoint.

Since for any pair of differentiable functions $d(t)$ and $l(t)$, with $l(t) > l_0$, for some $l_0>0$,  $t\mapsto H(l(t)) + K(l(t),d(t))$
is a one-parameter family of Schr\"{o}dinger operators on a common domain of self-adjointness $D$, by standard abstract methods (see e.g.\ Theorem~X.70 in~\cite{Reed}) the time-dependent  Schr\"odinger equation~(\ref{eq:StaticH_SchrEq}) is well defined for any initial condition $\phi(0)\in D$, and yields a unique unitary propagator.

Equation~(\ref{eq:StaticH_SchrEq}) is a central result of this paper. We now check its consequences on some interesting physical situations.

\subsection{Some Examples}

Consider the case in which there is only a translation of
the walls without dilation:
\begin{equation}
\dot{l}=0, \qquad d(t)=d_0+vt.
\end{equation}
The Schr\"odinger operator in the static reference frame, by assuming $l=1$, is
\begin{equation}
H+K= \frac{p^2}{2m}- v\, p
\end{equation}
which is exactly the Hamiltonian transformed according to the
Galilean transformation, as expected.

On the other hand, if we consider the case (constant acceleration $g$) 
\begin{equation}
\dot{l}=0, \qquad d(t)=d_0+\frac{1}{2}gt^2\,,
\end{equation}
we find:
\begin{equation}
H=\frac{p^2}{2m}-gtp,
\end{equation}
(taking again $l=1$).
In such a case, by  applying the gauge (unitary) transformation
\begin{equation}
G(t): L^2\left(-\frac{1}{2},\frac{1}{2}\right) \rightarrow
L^2\left(-\frac{1}{2},\frac{1}{2}\right) : \phi \mapsto \chi =
G(t)\phi\,,
\end{equation}
with
\begin{equation}
\chi(\xi) = (G(t)\phi)(\xi)=\e^{\frac{\ii}{\hbar}
\left(mg\xi t-\frac{1}{6}mg^2t^3\right)}\phi(\xi)\,,
\end{equation}
the domain $D$ is left invariant, $G(t)(D)=D$, and the
Hamiltonian becomes
\begin{equation}
G(t)H G^\dagger(t)=\frac{p^2}{2m}-\frac{1}{2}mg^2t^2.
\end{equation}
Moreover,
\begin{equation}
\ii \hbar \dot{G}(t)G^\dagger(t)=-mgx+\frac{1}{2}mg^2t^2,
\end{equation}
so that the new Schr\"odinger equation is
\begin{equation}
\ii \hbar \frac{\d}{\d t} \phi(t)=\left(\frac{p^2}{2m
}-mgx\right)\phi(t),
\end{equation}
and describes a particle in a constant gravitational field, in agreement with the equivalence principle. 
Notice, however, that the particle is in a box (a falling elevator).

\section{Energy rate equation}\label{sec:energy}

Let us consider an initial state $\phi(0)$ at time $t=0$. It will evolve into
the state $\phi(t)$
solution of the Schr\"odinger equation~(\ref{eq:StaticH_SchrEq}).

The energy of the system has a rate given by
\begin{equation}
\dot{E}(t)=\frac{\d }{\d t}\BraKet{\phi}{H(l)
\phi}=\frac{\ii}{\hbar}\big(\BraKet{\phi }{K H\phi}- \BraKet{K H
\phi}{\phi }\big)+\BraKet{\phi }{\dot{H} \phi},
\label{eq:ev_exp_val}
\end{equation}
with $K(l,d)$ given by~(\ref{eq:transport}), with domain
\begin{equation}
D(K)=\{f\in \mathcal{H}^1(I), \xi f'(\xi)\in
L^2(I)\},
\end{equation}
and $\phi \in D(KH)$, where
\begin{equation}
D(KH)=\left\{f\in \mathcal{H}^3(I),\;
f\left(-\frac{1}{2}\right)=f\left(\frac{1}{2}\right)=0\right\}.
\label{eq:D(KH)}
\end{equation}
A double integration by parts gives
\begin{eqnarray}
\frac{\d}{\d t}\BraKet{\phi(t)}{H(l(t)) \phi(t)} &=& -\frac{\hbar^2}{2 m
l^3}\left[\left( \dot{l}(t)\ \xi +\dot{d}(t) \right)
|\phi'(\xi,t)|^2 \right]_{-\frac{1}{2}}^{\frac{1}{2}} ,
\label{exp_val}
\end{eqnarray}
with $\phi'(\xi,t)= \partial_\xi \phi(\xi,t)$.

In the original frame, since  $\phi'(\xi,t)= l^{3/2} \psi'(l\xi+d,t)$, this yields
\begin{eqnarray}
\fl\quad\dot{E}(t)&=&\frac{\d}{\d t}\BraKet{\psi(t)}{H_0(l(t),d(t)) \psi(t)}
= -\frac{\hbar^2}{2 m}\left[\dot{b}(t) |\psi'(b(t) ,t)|^2 - \dot{a}(t) |\psi'(a(t) ,t)|^2 \right],
\label{eq:Erate}
\end{eqnarray}
with $a= -l/2+d$ and $b=+l/2+d$ the positions of the walls. As
expected, due to the movement of the walls, the energy change is due only to boundary terms depending on the velocities
of the walls. Indeed, within the walls the problem is that of a free particle, so that the energy changes only because of the presence of the moving walls. The case $d=\frac{l}{2}$ is the standard Fermi-Ulam accelerator model, where one of the two walls is still at $a=0$, while the other is at $b=l$. In such a case the rate~(\ref{eq:Erate}) particularizes to the result heuristically found by Konishi and Paffuti~\cite{Konishi}.

\subsection{Two Examples}

Let us now consider two examples that are valid within the limits of validity of perfectly reflecting walls.
Assume
\begin{equation}
l=l_0+vt, \qquad d=0.
\end{equation}
In this case the Schr\"odinger equation in the reference frame with fixed walls becomes
\begin{equation}
\ii \hbar \frac{\d }{\d
t}\phi=\frac{1}{l_0+vt}\left(\frac{1}{l_0+vt}\frac{p^2}{2m
}-v\,\xi \circ p\right)\phi,
\end{equation}
while the energy rate is
\begin{equation}
\dot{E}(t) =
-\frac{v}{(l_0+vt)^3}\frac{\hbar^2}
{2m}\frac{1}{2}\left[\left|\phi
'\left(\frac{1}{2},t\right)\right|^2+\left|\phi
'\left(-\frac{1}{2},t\right)\right|^2\right].
\end{equation}

As another example, we take
\begin{equation}
l=l_0+\sin(\omega t), \qquad d=0.
\end{equation}
The Schr\"odinger equation in the reference frame  is
\begin{equation}
\ii \hbar \frac{\d }{\d t}\phi=\frac{1}{l_0+\sin(\omega
t)}\left(\frac{1}{l_0+\sin(\omega t)}\frac{p^2}{2m
}+\omega\cos(\omega t) \, \xi \circ p\right)\phi,
\end{equation}
and the energy rate reads
\begin{equation}
\dot{E}(t) =
-\frac{\omega\cos(\omega t)}{(l_0+\sin(\omega
t))^3}\frac{\hbar^2}{2m} \frac{1}{2}\left[\left|\phi
'\left(\frac{1}{2},t\right)\right|^2+\left|\phi
'\left(-\frac{1}{2},t\right)\right|^2\right].
\end{equation}

\subsection{A Remark on  domains}

In the derivation of the energy rate given above we have been
somewhat cavalier about domains. Formally, the
expression~(\ref{eq:ev_exp_val}) is nothing but the expectation
value of the commutator $[K,H]$ (plus the
expectation of $\dot{H}$). However, as written there, that
expression is valid in a larger domain than $D([K,H])$, contained
in its form domain. We now show that such an extension is needed,
since for functions  in $D([K,H])$ the energy rate identically
vanishes.

Let us take a closer look at the domains by focussing on their
relevant part in $L^2(I)$. The Hamiltonian $H$ is defined on the
Dirichlet domain~(\ref{eq:Dirichlet}), $D(H)=D$, while (the
restriction of) $K$ is defined on $D(K)=\mathcal{H}^1(I)$,
so that the domain of the commutator  is
\begin{eqnarray}
\fl \quad D([K,H])&=&\left\{ \phi\in D(K), \; K \phi\in D(H)\right\}\cap\left\{ \phi\in D(H),\; H \phi\in D(K)\right\}
\nonumber \\
\fl \quad &=&\left\{\phi\in \mathcal{H}^3(I),\;
\phi\left(-\frac{1}{2}\right)=\phi\left(\frac{1}{2}\right)=0, \phi
'\left(-\frac{1}{2}\right)=\phi
'\left(\frac{1}{2}\right)=0\right\}.
\end{eqnarray}
As a consequence, for every $\phi\in D([K,H])$, one gets from~(\ref{exp_val})
\begin{equation}
\frac{\d}{\d t}\BraKet{\phi(t)}{H(l) \phi(t)}=0.
\end{equation}
This means that the space $D([K,H])$ is too small for our purposes
so we have to consider a larger space, such as $D(KH)$ in~(\ref{eq:D(KH)}),
which is contained in the form domain of the commutator.

\section{Adding a potential}\label{sec:potential}

In order to complete our treatment, we consider now the case in
which the particle is confined in a moving and dilating box, 
subject to the action of a time-dependent potential. In such a
situation, the Hamiltonian of the system is given by
\begin{equation}
H_1(l,d,t)=\left(-\frac{\hbar^2}{2 m} \frac{\d^2}{\d x^2} + V(x,t) \right)
\oplus_{l,d} 0\, , \label{Ham_potential}
\end{equation}
on the domain $D_{l,d} \oplus L^2(I_{l,d}^c)$. We assume that
$V(\cdot,t):I_{l,d}\to\mathbb{R}$ is a measurable function such
that $H_1(l,d,t)$ is self-adjoint. For this purpose it suffices,
e.g., that $V(\cdot,t) \in L^{2}(I_{l,d})$ for every $t$, $l$, and $d$.

Under the action of the unitary transformation $U$, the potential
is transformed in a simple way:
\begin{equation}
\tilde{V}(\cdot,t) = U(l,d) V(\cdot,t) U^\dag(l,d) \,, \qquad  \tilde{V}(\xi,t) = V(l\xi+d,t)\,.
\end{equation}
The rate equation for the energy of the particle is given by
\begin{eqnarray}
\nonumber
\fl\qquad \dot{E}_1(t)=\frac{\d }{\d t}\BraKet{\psi(t)}{H_1(l,d,t)\psi(t)}
&=&\frac{\ii}{\hbar}\big(\BraKet{\phi }{K H\phi}- \BraKet{K
H \phi}{\phi }\big)+\BraKet{\phi }{\dot{H} \phi} \\
\fl \qquad &+& \frac{\ii}{\hbar}\big(\BraKet{\phi }{K \tilde{V}\phi}- \BraKet{K \tilde{V}
\phi}{\phi }\big)+\BraKet{\phi }{\dot{\tilde{V}} \phi}\,.
\label{eq:ev_exp_val_2}
\end{eqnarray}
Since $\dot{\tilde{V}} = -\ii/\hbar[K,\tilde{V}]+U \partial V /
\partial t U^\dag$, and considering Eq.~(\ref{eq:ev_exp_val}), one gets
\begin{eqnarray}
\dot{E}_1(t)= \dot{E}(t) + \Bra{\phi} U \frac{\partial V}{\partial t} U^\dag \Ket{\phi}
\,. \label{eq:Rate_With_Potential}
\end{eqnarray}

It is worth noting that while a detailed evaluation of the mean
value of $\ii/\hbar[K,H]+\dot{H}$ gives rise to contact terms, in
the case of $\ii/\hbar[K,\tilde{V}]+\dot{\tilde{V}}$ no similar contributions appear. This is due to the fact that $H$ is a differential operator, while $\tilde{V}$ is simply a
multiplication operator.

As a special case, let us consider the situation where $V$ is
essentially a static potential, meaning that there exists a
certain $\varphi\in L^2_{\mathrm{loc}}(\mathbb{R})$, independent of time, such that $V(x,t) =
\varphi(x)$ for $x\in I_{l,d}$ and $V(x,t)=0$ elsewhere.
In this case it turns out that the presence of the potential does not add extra terms to
the energy variation rate. Formally, time
derivatives of $V$ can produce $\delta$-functions at the border,
but since the wave function satisfies Dirichlet boundary
conditions, the net result of the integration in
Eq.~(\ref{eq:Rate_With_Potential}) is zero.

\section{Conclusions}\label{sec:conclusion}

We have analyzed the problem of the non-relativistic quantum bouncer. 
Our focus has been on
the very structure of the problem and on the possible difficulties of framing it in a 
rigorous mathematical framework.
The first question that we have
tackled is the correct formulation of the problem in mathematical
terms, avoiding the use of equations defined in different Hilbert
spaces at different times. To this end, we have used an extended
Hilbert space and defined all the operators as zero outside the box
and non-zero (with their standard action) inside the box.
Subsequently, to effectively solve the relevant dynamical problem,
we have exploited a unitary transformation that converts the
original problem into the problem of a particle moving in a fixed
box but governed by a time-dependent Hamiltonian. This has required a proper redefinition of the problem, from that of a system evolving under time-dependent boundary conditions into that of a system suitably evolving
under fixed boundary conditions.
In such a way, we have ended up with a family of time-dependent Hamiltonian operators with a common
domain of self-adjointness. As a consequence, the time-dependent Schr\"{o}dinger
equation is well-defined for any initial vector in the common domain, and
the existence of a unique unitary propagator is assured.

An interesting aspect of our analysis is the possibility to write
down a general rate equation describing the energy change
undergone by the particle. In fact, also from a classical point of
view, a particle interacting with moving walls is subject to
some variation of kinetic energy, and this has a quantum mechanical
counterpart, well expressed by the contact terms in
Eq.~(\ref{exp_val}). Moreover, we provided a further development by
considering the effects of a potential. The
relevant new terms in the rate equation of the energy do not
include contact terms.

A possible extension of
our results concerns the analysis of different boundary conditions and/or boxes in two- and
three-dimensional spaces.

\appendix

\section{Building up the walls by  a product formula}\label{app:Zeno}
\label{sec:appendix}

We derive here the Schr\"odinger equation for the static
frame, starting from the Schr\"odinger equation of a free particle
on a line and constructing the moving walls as the limit of
infinitely time-dependent spatial projections, a limit known as
quantum Zeno dynamics~\cite{ZenoMP}.
Consider the projections
\begin{equation}
P(t) = \chi_{\left(-l(t)/2 +d(t),l(t)/2+d(t)\right)}(x),
\end{equation}
$\chi_A$ being the characteristic function of the set $A$. Then
$P(t)=U^\dagger PU$, with
$P=\chi_{\left(-\frac{1}{2},\frac{1}{2}\right)}$ the
projection on the moving box. If we discretize  time: $t_k= kt/N$
where $k=0,\dots,N$, the evolution in the moving box is
approximated by the operator ($\hbar=1$):
\begin{eqnarray}
W_N(t)&=&P(t_N)\e^{-\frac{\ii  Tt}{N}}\cdots \e^{-\frac{\ii Tt}{N}}P(t_1)\e^{-\frac{\ii Tt}{N}}P(t_0)\\
\nonumber
&=&U^\dagger(t_N)\prod _{k=1}^{N}[PU(t_k)\e^{-\frac{\ii Tt}{N}}U^\dagger(t_{k-1}) P]U(0),
\end{eqnarray}
where $T=\frac{p^2}{2m}$ is the kinetic energy operator  on the
whole line $\mathbb{R}$ and we used the notation
$U(t_k)=U(l(t_k),d(t_k))$. Now we formally compute~\cite{Zeno unchained}
\begin{eqnarray}
\fl \qquad PU(t_k)\e^{-\frac{\ii Tt}{N}}U^\dagger(t_{k-1})P&=& \exp\left(-\frac{\ii t}{N}P\left(T(t_{k-1})+\ii\dot{U}(t_{k-1})U^\dagger(t_{k-1})\right)P\right) P \nonumber\\
& & +O\!\left(\frac{t^2}{N^2}\right) ,
\end{eqnarray}
So that, in the limit $N\to \infty$
\begin{equation}
W(t)=\lim_{N\to \infty}W_N(t)=U^\dagger(t)\mathcal{T}\e^{-\ii \int_{0}^{t}P\left(T(s)+\ii\dot{U}(s)U^\dagger(s)\right)Pds}P U(0),
\end{equation}
$\mathcal{T}$ denoting time-ordering.
The evolution in the original frame is therefore
\begin{equation}
\psi(t)=W(t)\psi(0),
\end{equation}
while in the static reference frame 
\begin{equation}
\phi(t)=U(t)W(t)U^\dagger(0)\phi(0)=\mathcal{T}\e^{-\ii \int_{0}^{t}P\left(T(s)+\ii\dot{U}(s)U^\dagger(s)\right)Pds}P\phi(0).
\end{equation}
In conclusion, the Schr\"odinger equation in the frame with fixed box
reads
\begin{equation}
\ii  \dot{\phi}(t)=(H(t)+\ii \dot{U}(t)U^\dagger(t))\phi(t),
\label{eq:newSchrodinger1}
\end{equation}
where $H$ is the operator defined in Eq.~(\ref{eq:H(l,d)}).

\end{document}